\documentstyle[12pt]{article}
\newcommand{\be}{\begin{equation}}
\newcommand{\ee}{\end{equation}}
\newcommand{\ds}{\displaystyle}
\renewcommand{\theequation}{\arabic{section}.\arabic{equation}}
\begin{document}
\mbox{ }\hfill{\normalsize ITP-94-19E}\\
\mbox{ }\hfill{\normalsize April 1994}\\
\begin{center}
{\sloppy \LARGE Polarization observables of the
$\vec{d} \vec{p} \rightarrow \vec{p}d$ reaction
and  one-neutron-exchange approximation}\\
\vspace{0.5in}

{A.~P.~Kobushkin and A.~I.~Syamtomov \\
  {\it N N Bogulyubov Institute for Theoretical Physics} \\
  {\it 252143 Kiev, Ukraine} \\
         C.~F.~Perdrisat \\
  {\it the College of William and Mary} \\
  {\it Williamsburg, VA 23185, USA} \\
  and \\
  V.~Punjabi \\
  {\it Norfolk State University} \\
  {\it Norfolk, VA, 23505, USA}}
\end{center}

\begin{abstract}

The polarization observables in the elastic scattering of polarized deuterons
on a polarized hydrogen target, with measurement of the recoil
proton polarization, are considered. The observables are calculated in the
one-neutron exchange approximation, for the special case of backward
scattering
($\theta_{c.m.} = 180^{\circ}$). Several new relations between polarization
observables of the reaction are derived within the framework of this
approximation.
\end{abstract}
\newpage

\section{Introduction}
\setcounter{equation}{0}

    One of the most interesting problems of contemporary nuclear physics is
to define the transition region from the traditional, purely hadronic
description of nuclear structure, to a quark-gluon
description. Hadron induced reactions in the energy range of a few GeV's
play an important role in bringing light on this problem.
Of particular interest are relativisitic, polarized beams of deuterons,
because the spin 1 of the deuteron allows
for a large number of polarization experiments; several vector and tensor
polarization transfer coefficients can be
measured besides the usual analyzing powers.

        Results from recent measurements
with polarized deuteron beams in Dubna and Saclay have
started exploiting this potential. The tensor analyzing power, $T_{20}$,
in inclusive breakup at $0^\circ$ was measured
on several nuclei including
Hydrogen by Perdrisat et al \cite{perdr87};
and with a  Carbon target at
9 $GeV/c$ by Ableev et al \cite{ablt20}. Similarly,
the polarization transfer, $\kappa_0$, was obtained
for Hydrogen by Cheung et al \cite{cheung}, and
for Carbon by Sitnik et al \cite{sitnik} and Strunov et al \cite{kappa}.
The first
measurement of $T_{20}$ in backward elastic $dp$ is that of Arvieux et al
\cite{arvieux}; recently these measurements have been repeated
by Perdrisat et al \cite{saturne} up to 3.6 $GeV/c$, and extended to
6.0 $GeV/c$ by Piskunov et al \cite{T20dubna}. In addition the polarization
transfer for backward elastic scattering was obtained by Punjabi et al
\cite{punjabi}.

        An important point has been established on the basis
of the analysis of the cross section data for the inclusive
$(d,p)$ breakup process at $0^\circ$ and backward elastic dp.
Kobushkin \cite{kobush86} has shown that the cross sections of
these two reactions could be reproduced with a common ``empirical
momentum distribution''. This was done using the one-neutron-exchange (ONE)
model, which is a particular version of the IA equivalent to the spectator
model.
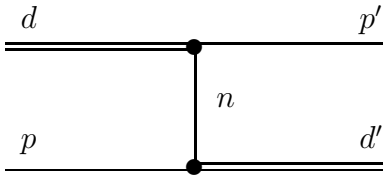
\begin{figure}[b]
\setlength{\unitlength}{0.1cm}
\begin{picture}(100,25)
\tenrm
\put(45,22){\rule[-.15mm]{5.cm}{.25mm}}
\put(45,21){\rule{2.5cm}{.25mm}}
\put(58,22.5){\line(4,-1){2}}
\put(58,20.6){\line(4,1){2}}
\put(83,22.5){\line(4,-1){2}}
\put(83,21.5){\line(4,1){2}}
\put(70,21.5){\circle*{2}}
\put(70,5.5){\circle*{2}}
\put(45,5){\rule{5.cm}{.25mm}}
\put(70,6){\rule[-.15mm]{2.5cm}{.25mm}}
\put(83,6.5){\line(4,-1){2}}
\put(83,4.7){\line(4,1){2}}
\put(58,5.7){\line(4,-1){2}}
\put(58,4.7){\line(4,1){2}}
\put(70,6){\rule{.25mm}{1.5cm}}
\put(69.5,14){\line(1,-4){0.5}}
\put(70.3,12){\line(1,4){0.5}}
\put(47,24){$d$}
\put(92,24){$p^{\prime}$}
\put(47,8){$p$}
\put(92,8){$d^{\prime}$}
\put(73,13.5){$n$}
\end{picture}
\caption{The one-neutron-exchange diagram for
the elastic $pd$ backward scattering.}
\end{figure}
It shows that these two reactions are very closely related, as could
be inferred from the similarity of their Feynman graphs. Therefore, these
two reactions bring complementary information about the deuteron structure.
Of course it can not be excluded {\it a priori} that in such
reactions the ONE contribution is masked by other mechanisms such as pion
exchange and -or pion production, as discussed by Lykasov \cite{lykasov},
delta excitation in the intermediate state and so on. There is no clean way of
separating these different reaction mechanisms; yet, as discussed
recently by Kuehn, Perdrisat and Strokovsky \cite{strok}, there are empirical
indications that the relativistic form
of ONE used by Kobushkin \cite{kobush93} is a good approximation.

        The goal of this paper is to discuss measurements of
additional polarization observables.
The paper was inspired by the talk of Ladygin, Rekalo and Sitnik
\cite{Ladygin}, in which they considered measurements of the asymmetry
in backward elastic
scattering of polarized deuterons from a polarized hydrogen target.
The remainder of this paper is organized  as  follows. In Sec.2 observables
of the $\vec{d} \vec{p} \rightarrow \vec{p}d$-reaction are discussed
without reference to a particular reaction mechanism. Then in
Sec.3 we calculate the corresponding polarization observables for the
scattering angle $\theta_{c.m.} = 180^{\circ}$ in the ONE-approximation.
Within  the
specified assumptions a set of relations between the polarization
observables of the reaction is derived. Finally, the results are
summarized in Sec.4.

\section{Polarization observables}
\setcounter{equation}{0}

In what follows we will use the upper indices for the polarization
of the final particles and the lower ones for that of the initial
particles.
We will consider only the case where
the beam and the target quantization axes are
parallel and both are perpendicular
to the incident deuteron beam direction. Following the common
practice\cite{olsen}, we refer the polarization observables to the
projectile helicity frame, where the +$z$-axis is directed along the beam
direction ${\bf k}_{in}$ and $y$ is taken along
${\bf k}_{in}\times {\bf k}_{out}$; ${\bf k}_{out}$ represents
the momentum direction of the scattered particle. Finally, the $x$-
direction is chosen to form a right-handed coordinate system. The direction
of ${\bf k}_{in}\times {\bf k}_{out}$ is undefined for the
case of $\theta_{c.m}=180^\circ$; we choose the $y$-axis directed along the
quantization axis of the deuteron beam.

If $N_{+}$, $N_{-}$ and
$N_{0}$ - are the occupation numbers of the deuteron
substates with the spin projections $+1$, $-1$ and $0$, respectively, and
$n_{\pm}$ are the occupation numbers of the target proton substates with
the spin projections $\pm \frac{1}{2}$, then the deuteron vector polarization
$p_y$, its tensor polarization $p_{yy}$ and the proton polarization
$\mu_y$ are defined as
\be
p_{y}=N_{+}-N_{-} ,\ \ p_{yy} =N_{+}+N_{-}-2N_{0},\\
\label{2.1}
\ee
\be
\mu_{y}=n_{+}-n_{-},
\label{2.2}
\ee
(where, by definition, $N_{+}+N_{-}+N_{0} = 1$ and
$n_{+}+n_{-} = 1$). Considering $p_y,\ p_{yy}$ and $\mu_{y}$
as initial parameters of the reaction, from (\ref{2.1}) and (\ref{2.2})
one gets
\be
N_{\pm} = \frac{1}{3} \pm \frac{1}{2}p_y+\frac{1}{6}
p_{yy}, \ \ \ N_{0} = \frac{1}{3}(1 - p_{yy}),
\label{2.3}
\ee
\be
n_{\pm} = \frac{1}{2}(1 \pm \mu_y) .
\label{2.4}
\ee
The differential cross section for the scattering of a polarized deuteron on a
polarized proton target, when the recoil proton has the spin
projection $m^{\prime}$, is then
\be d\sigma^{m^{\prime}} =
\sum_{Mm}N_{M}n_{m}d\sigma^{m^{\prime}}_{Mm} =
\sum_{MM^{\prime}m}N_{M}n_{m}d\sigma^{M^{\prime}m^{\prime}}_{Mm} ,
\label{2.5}
\ee
where $M$, $M'$ and $m$  are the spin projections of the incident and outgoing
deuterons, and of the target proton, respectively, and
$d\sigma^{M^{\prime}m^{\prime}}_{Mm}$ is the differential
cross section for  $dp$ elastic scattering for  pure spin substates.
 Assuming first that the polarization of the final particles is not measured,
expression (\ref{2.5}), after  taking into account
(\ref{2.3}) and (\ref{2.4}), is reduced to the following
\begin{equation}
d\sigma = d\sigma_{\mbox{nonpol}}(1 - \frac{1}{4}p_{yy}A_{zz} +
\frac{3}{2}p_y\mu_yC_{y,y}) ,
\label{2.6}
\end{equation}
where
\begin{equation}
d\sigma_{\mbox{nonpol}} =\frac{1}{3}( d\sigma_{1\ \frac{1}{2}}+d\sigma_{-1\
\frac{1}{2}}+d\sigma_{0\ \frac{1}{2}}) ,
\label{2.7}
\end{equation}
\begin{equation}
A_{zz} = -2\frac{ d\sigma_{1\ \frac{1}{2}}+
d\sigma_{-1\ \frac{1}{2}} - 2 d\sigma_{0\ \frac{1}{2}}}
{ d\sigma_{1-\frac{1}{2}} + d\sigma_{1-\frac{1}{2}} + d\sigma_{0-
\frac{1}{2}}} ,
\label{2.8}
\end{equation}
\begin{equation}
C_{y,y}= \frac{d\sigma_{1\frac{1}{2}} - d\sigma_{1-\frac{1}{2}}}
{ d\sigma_{1\frac{1}{2}} + d\sigma_{1-\frac{1}{2}} + d\sigma_{0
\frac{1}{2}}} ,
\label{2.9}
\end{equation}
are the cross section for unpolarized particles, the tensor analyzing power
(the spherical notation $T_{20}\equiv \frac{1}{\sqrt2}A_{zz}$ is usually
used) and the spin correlation coefficient of the reaction
(see Ohlsen \cite{olsen}), respectively, and
$d\sigma_{Mm} =  \sum_{m^{\prime}}{} d\sigma_{Mm}^{m^{\prime}}$.
For the final proton polarization
\be
\mu^y =  \frac{ d\sigma^{+\frac{1}{2}} -d\sigma^{-\frac{1}{2}}}
{ d\sigma^{+\frac{1}{2}} + d\sigma^{-\frac{1}{2}}} ,
\label{2.10}
\ee
one has
\be
\mu^y\ d\sigma  = d\sigma _{\mbox{nonpol}}\left(\mu_yK^{0,y}_{0,y}-
\frac{1}{4}p_{yy} \mu_y K^{0,y}_{zz,y} + \frac{3}{2}p_y K_{y,0}^{0,y}\right).
\label{2.11}
\ee
The physical meaning of the coefficients
$K_{y,0}^{0,y}$, $K^{0,y}_{0,y}$ and $K^{0,y}_{zz,y}$
in the expression (\ref{2.11}) is the following:
$K_{y,0}^{0,y}$   is the transfer coefficient of the polarization from a
vector-polarized deuteron to the recoil proton \cite{olsen},
when both the tensor polarization of the incident deuteron and the target
polarization are zero; and $K^{0,y}_{0,y}$ and $K^{0,y}_{zz,y}$ are the
vector and tensor coefficients for the recoil proton polarization,
for unpolarized and tensor polarized
deuteron beam, respectively, but polarized proton target. The
quantity $\kappa_0=\frac{\ds 3}{\ds 2}K_{y,0}^{0,y}$ is used almost
universally as the polarization transfer coefficient.
\section{\sloppy
Polarization ob\-ser\-va\-bles in ONE-ap\-pro\-xi\-ma\-tion}
\setcounter{equation}{0}

The non-relativistic wave function of the deuteron with  spin
projection $M$ is expressed via the $S$- and $D$-radial wave functions
$u(k)$ and $w(k)$ as
\begin{eqnarray}
\lefteqn{| M \rangle = \left[\sqrt{\frac{1}{4\pi}} u(k) \sum_{m, m^{\prime}}
      \langle \frac{1}{2} \frac{1}{2} m m^{\prime}|1 M \rangle \right. +}
      \nonumber \\
& + & \left.  w(k) \sum_{m, m^{\prime}, \sigma, M^{\prime}}
      \langle \frac{1}{2} \frac{1}{2} m m^{\prime}|1 M^{\prime} \rangle
      \langle 2 1 \sigma M^{\prime}|1 M \rangle Y^{\sigma}_{2}(\hat{k})
      \right]\chi_{m}^{p}\otimes \chi_{m^{\prime}}^{n} ,
\label{3.1}
\end{eqnarray}
where $\langle j j^{\prime} m m^{\prime}|J M \rangle$ are Clebsh-Gordan
coefficients, $\chi_{m}^{p}$ and $\chi_{m}^{n}$ are  the proton and neutron
spinors, respectively, and $\vec k$ is the ``internal'' momentum in the
deuteron.
The latter is defined in the non-relativistic theory in terms of the
momenta of the deuteron $\vec d$ and of the proton-spectator $\vec p$ , both
in the Center of Mass System (CMS), as $\vec k = \frac{1}{2} \vec d - \vec p$.
It is commonly accepted (see e.g. Frankfurt and Strikman \cite{strikfra}),
that in the light front dynamics (LFD) the relativistic deuteron wave
function has the same spin structure
as in (\ref{3.1}), but the internal momentum $\vec k$ is connected to the
momenta $\vec d$ and $\vec p$ in a more complicated way. For the reaction
under study here the corresponding expression can be found in
\cite{kobush86}.
Thus, in the ONE-approximation the reaction amplitude is:
\begin{equation} T_{MmM^{\prime}m^{\prime}}(\vec k) = t(k)
 \langle M^{\prime}|\chi_{m}^{p} \chi_{m'}^{p' \dagger}| M \rangle ,
\label{3.2}
\end{equation}
where $t(k)$ is the part of the scattering amplitude which is independent of
the spin variables, and $M$ and $M^{\prime}$  are the initial and final
deuteron spin projections, respectively;  $m$ and $m^{\prime}$ are the
spin projections for the target and
recoil proton, respectively. Here it should be emphasized that the spinors
$\chi_{m}^{p}$ and $\chi_{m^{\prime}}^{p^{\prime}}$ have to do with
different protons, and therefore with different spin subspaces.

Using an explicit expressions for the amplitudes
$T_{MmM^{\prime}m^{\prime}}(\vec k)$ (see Appendix), one gets the following
formulae for the polarization observables
\begin{eqnarray}
A_{zz} & = & \frac{2\sqrt{2}uw-w^2}{u^2+w^2} ,
\label{3.3} \\
C_{y,y} & = &
\frac{2}{9} \frac{u^4-2w^4+3u^2w^2-uw(5u^2-2w^2)/ \sqrt{2}}{(u^2+w^2)^2},
\label{3.7} \\
K_{0,y}^{0,y} & = &
\frac{1}{9}\left[  \frac{(u-\sqrt{2}w)^2}{u^2+w^2} \right]^2 ,
\label{3.5} \\
K_{zz,y}^{0,y} & = &
-\frac{8}{9} \frac{(u-\sqrt{2}w)^2(u^2+\frac{7}{8}w^2+uw/(2\sqrt{2}))}
 {(u^2+w^2)^2},
\label{3.6} \\
\kappa_{0} & = &
\frac{u^2-w^2-\sqrt{\frac{1}{2}}uw}{u^2+w^2} .
\label{3.4}
\end{eqnarray}
The expressions (\ref{3.3}) and (\ref{3.4})  were first
 obtained for elastic scattering by Vasan
\cite{Karman1} and Ladygin et al \cite{Ladygin1}, respectively.
Expression (\ref{3.7}) has been first shown in Ref.\cite{Ladygin}.
In the ONE-approximation
$T_{20}(k)$ and $\kappa_{0}(k)$ of the
$\theta_{c.m.} = 180^{\circ}$ elastic scattering reaction
are identical with $T_{20}(k)$ and $\kappa_{0}(k)$ of the deuteron breakup
reaction $(d,p)$, with the proton detected at $0^{\circ}$, calculated
in the impulse approximation. It was  pointed out earlier
\cite{strok} that the functions
$T_{20}(k)$ and $\kappa_{0}(k)$ obtained in the ONE-approximation
(for the elastic $dp$-scattering) or in the impulse approximation
(for the $(d,p)$-reaction), are related through the simple identity
\begin{equation}
\left(T_{20} +\frac{1}{2\sqrt2} \right) ^2 + \kappa_0^2 = \frac{9}{8}.
\label{3.9}
\end{equation}
Using (\ref{3.3})-(\ref{3.4}) one can easily obtain an another set of
relations between the polarization observables of the reaction under
consideration
\begin{eqnarray}
K_{0,y}^{0,y} & = & \frac{1}{9}(1 - \sqrt2 T_{20})^2,
\label{3.10} \\
K_{zz,y}^{0,y} & = & -\frac{8}{9}(1 - \sqrt2 T_{20}) \left(1
+\frac{1}{4\sqrt2}T_{20} \right)
\nonumber \\
& &\hspace{0.5in} = -\sqrt{\frac{K_{0,y}^{0,y}}{2}} \left(3 -
\sqrt{K_{0,y}^{0,y}}\right),
\label{3.11} \\
C_{y,y} & = &
\frac{2}{9} \kappa_0 (1 - \sqrt2 T_{20})=\frac{2}{3}\kappa _{0}
\sqrt{K_{0,y}^{0,y}}.
\label{3.12}
\end{eqnarray}
It is important to emphasize that the above relations, as well as the
``circle'' relation (\ref{3.9}), do not depend on a particular choice
of the deuteron wave functions $u(k)$ and $w(k)$. Their validity is based
upon only two assumptions: the ONE-approximation applicability and the
hypothesis \cite{strikfra}  that in the LFD the deuteron
wave function has only two ($S$- and \mbox{$D$-)} angular momentum components.
 In the general case the relativistic deuteron wave function
may have  additional components, besides the usual $S$- and $D$- components
(see e.g. Gross \cite{gross} and Karmanov \cite{karman}). Therefore, the
experimental characterization of the relations derived here
must throw  additional light upon the structure of the
relativistic deuteron at short distances.

\section{Conclusions and discussion}
\setcounter{equation}{0}

In the present paper we considered the polarization observables
of the elastic scattering of a polarized deuteron on a polarized proton
at $\theta_{c.m.} = 180^{\circ}$, when the recoil proton polarization is
measured.
A set of four relations between the five polarization observables has
been derived within the framework of the ONE-approximation (fig.1) and
assuming that in the LFD the deuteron relativistic wave
function possess only $S$- and $D$- components.
As an illustration of these relations we show in figs.2,3 the
dependencies of the spin correlation
coefficient $C_{y,y}$ and the
vector coefficient of the recoil proton polarization $K_{0,y}^{0,y}$ upon
$T_{20}(k)$ and $\kappa_{0}(k)$. If both the ONE and LFD hypotheses
were fully verified, the corresponding observables, depending on the internal
momentum $k$, would stay on the curves  which are at the intersection of the
surface
(\ref{3.12}) or (\ref{3.10}) with the cylinder built upon the circle
(\ref{3.9}) in the plane $T_{20}(k)$--$\kappa_{0}(k)$.

In addition six new observables could be obtained by measuring the final
state deuteron polarization, but these are constrained by six relationships
similar to (\ref{3.9})-(\ref{3.12}), also based on ONE.
The data for $T_{20}$ and $\kappa_0$ show that the first relation (\ref{3.9})
is violated in the case of inclusive breakup at $0^\circ$, as discussed in
\cite{strok}. Therefore, an experimental determination of all five
polarization
observables would provide us with a wealth of new information on the structure
of the deuteron.

 The authors express their thanks to V.~Ladygin and E.~Strokovsky for
interest in this work and stimulating discussions.

\section*{Appendix}
\setcounter{equation}{0}
\renewcommand{\theequation}{A.\arabic{equation}}

In the case of the scattering at $\theta_{c.m.} = 180^{\circ}$,
the spin amplitudes (\ref{3.2}) are
\begin{eqnarray}
T_{1m1m^{\prime}} & = &
 \frac{t(k)}{4\pi} \left[\left(u-\sqrt{{1\over 8}}w\right)^{2}
 \delta_{m\frac{1}{2}} \delta_{m^{\prime}\frac{1}{2}}
 + {9\over 8}w^{2}
 \delta_{m-\frac{1}{2}} \delta_{m^{\prime}-\frac{1}{2}}
 \right] , \\
\label{P.1}
T_{1m0m^{\prime}} & = &
 \frac{t(k)}{4\pi} \left[\sqrt{1\over 2}\left(u+\sqrt{{1\over 2}}w\right)
 \left(u-\sqrt{{1\over 8}}w\right)
 \delta_{m-\frac{1}{2}} \delta_{m^{\prime}\frac{1}{2}}+ \right.\nonumber\\
&  & \hspace{2.5cm} +\left.
{3\over 4} e^{2i\phi} \left(u+\sqrt{{1\over 2}}w\right) w
\ \delta_{m\frac{1}{2}} \delta_{m^{\prime}-\frac{1}{2}}
 \right] , \\
\label{P.2}
T_{1m-1m^{\prime}} & = &
 \frac{t(k)}{4\pi} \frac{3}{\sqrt{8}}\left(u-\sqrt{{1\over 8}}w\right)
 w\ \delta_{m m^{\prime}} ,\\
\label{P.3}
T_{0m0m^{\prime}}  & = &
 \frac{t(k)}{8\pi} \left(u+\sqrt{{1\over 2}}w\right)^{2}
 \delta_{m m^{\prime}} .
\label{P.4}
\end{eqnarray}

Taking into account
$|T_{MmM^{\prime}m^{\prime}}| =|T_{-M-m-M^{\prime}-m^{\prime}}| =
|T_{M^{\prime}m^{\prime}Mm}|$, one immediately gets the observables
above from these amplitudes. The polar angle
$\phi$ do not enter the final expressions for the cross section and the
polarization observables because of the choice $\theta_{c.m.} = 180^{\circ}$.

\newpage
\begin{center}
{\large Figure captions}
\end{center}

Figure 2. \parbox[t]{4.0in}{The curve at the intersection of the surface
(\ref{3.12})
with the cylinder built upon the circle (\ref{3.9}) in the plane
$T_{20}(k)$--$\kappa_{0}(k)$ describes the spin correlation
coefficient $C_{y,y}$ in the ONE-approximation.}
\vspace{0.2in}

Figure 3. \parbox[t]{4.0in}{The curve at the intersection of the surface
(\ref{3.10}) with the cylinder built upon the circle (\ref{3.9}) in the plane
$T_{20}(k)$--$\kappa_{0}(k)$ describes the
vector coefficient of the recoil proton polarization $K_{0,y}^{0,y}$
in the ONE-approximation.}

\end{document}